# Projection of global daily $CO_2$ emissions in 2026

Tao Li[1], Zhu Deng[2,*], Philippe Ciais[3], Steve J. Davis[4], Ling Ye[5], and Zhu Liu[1,*]

[1] Department of Earth System Science, Tsinghua University, Beijing 100084, China
[2] Department of Geography, The University of Hong Kong, Hong Kong 999077, China
[3] Laboratoire des Sciences du Climat et de l'Environnement LSCE, Gif-sur-Yvette 91191, France
[4] Department of Earth System Science, Stanford University, Stanford 94305, USA
[5] College of Geo-Exploration Science and Technology, Jilin University, Changchun 130026,China

*** Corresponding authors** Zhu Deng (zhudeng@hku.hk) and Zhu Liu (zhuliu@tsinghua.edu.cn).

## Abstract

Global fossil-fuel combustion and industrial-process $CO_2$ emissions remain close to historical highs, yet the recent slowdown in growth is difficult to assess before conventional inventories become available. We project emissions for 2026 and examine daily departures during major climate and energy disruptions. We fit Prophet models to the latest five years of Carbon Monitor activity estimates for 47 region-sector series. We use these models to project daily emissions for October–December and combine them with activity estimates through September. Global emissions are estimated at 38.0 Gt $CO_2$ in 2026, 34.1 Mt $CO_2$ (−0.1%) below 2025. The hindcast-adjusted range for the annual change is −0.9% to +0.7%. Five September-cutoff hindcasts show lower mean daily errors and larger mean absolute annual biases for Prophet than for a seasonal-naive benchmark. Regional increases total 0.331 Gt and decreases total 0.365 Gt, so 95.1% of the 0.696 Gt gross movement cancels in the global sum. China contributes the largest reduction (−0.243 Gt), concentrated in industry and ground transport, whereas India contributes the largest increase (+0.222 Gt), led by power and industry. Heat-period power anomalies differ among countries and can be muted by summer averaging, while aviation and shipping show different disruption and recovery profiles. Resolving the near-zero global change reveals where reductions are concentrated and where continued growth offsets them, providing a timely basis for assessing the breadth of mitigation progress.

# 1. Introduction

Global fossil-fuel combustion and industrial-process $CO_2$ emissions remain close to historical highs, although annual growth has slowed [1, 2]. Before complete annual inventories become available, it is difficult to determine whether this slowdown will persist as a high-emissions plateau or develop into a sustained decline. IEA projects continued growth in electricity demand from industry, cooling, and data centres [3, 4] alongside expanding renewable power generation and electrification [5]. Timely emissions estimates are therefore needed to assess mitigation progress and inform near-term policy [6, 7]. Assessing 2026 is more difficult because developing El Niño conditions and disruptions to Middle Eastern energy and transport markets coincided with changes in energy demand, mobility, and economic activity [8–14].

Conventional inventories and gridded emissions datasets provide national and global $CO_2$ estimates, though compilation and review delay their assessment of current developments [15–18]. Provincial emissions accounts provide subnational detail for China [19]. The Global Carbon Budget also tracks land-use emissions and the land and ocean sinks [20–23]. Near-real-time estimates draw on frequently updated activity indicators and sector-specific emission factors [24–26]. Daily $CO_2$ reconstructions extend the historical record back to 1970 [27], and daily plant-level estimates also cover air pollutants from power generation [28]. Carbon Monitor provides daily estimates for power generation, industry, ground transport, residential fuel use, aviation and shipping [29]. Activity-based estimates tracked the decline and uneven recovery in $CO_2$ emissions during COVID-19 [30–32], with satellite-based evidence for China [33]. Combining satellite retrievals can also help track fossil $CO_2$ emissions [34] and resolve fine-scale atmospheric $CO_2$ patterns over China [35]. Daily resolution helps distinguish brief departures from persistent changes that may cancel in an annual global total [36]. The full-year outcome remains uncertain while part of the year still needs to be projected.

Deng et al. [1] addressed this temporal gap by fitting additive Prophet models to 47 region-sector series and projecting each series from the activity-data cutoff to the end of the year. The approach extends near-real-time monitoring to a full-year outlook, allowing the annual total and its regional and sectoral components to be assessed before complete activity data become available.

We fit forecasting models to daily activity estimates through 30 September 2026 and project emissions for the final 92 days of the year across the 47 region-sector series. We first compare annual Carbon Monitor estimates with other inventories. We then decompose the 2026 global change by region, sector and period and assess its sensitivity using five historical hindcasts and a seasonal-naive benchmark. We examine daily departures during selected climate and energy windows, including national power-sector diagnostics for six countries.

# 2. Methodology and materials

## *2.1 Daily emissions and analytical scope*

We use daily activity-based estimates of fossil-fuel combustion and industrial-process $CO_2$ emissions from the September 2026 Carbon Monitor release [29]. Emissions for region $r$, sector $s$, and day $t$ are estimated from activity data or proxies ($AD$) and the corresponding emission factor ($EF$).

$$E_{r,s,t} = AD_{r,s,t} EF_{r,s}.$$

The analysis covers daily estimates from 1 January 2019 to 30 September 2026 for China, the United States, EU27, India, the Russian Federation, Japan, Brazil, the United Kingdom, and the rest of world (ROW). Country-level power diagnostics are evaluated independently in Section 2.5. Power, industry, ground transport, residential emissions, and domestic aviation are represented in each region, and the two global international aviation and shipping series are displayed under ROW. These form 47 non-overlapping accounting series. Residential and other fuel-use emissions cover direct use in IPCC 1.A.4–5. Emissions from electricity generation are allocated to power.

## 2.2 Daily projection

Following Deng et al. [1], we fit an additive Prophet model to each of the 47 region-sector series [37].

$$y_{r,s}(t) = g_{r,s}(t) + s_{r,s}^{year}(t) + s_{r,s}^{week}(t) + h_{r,s}(t) + w_{r,s}(t) + \varepsilon_{r,s}(t),$$

The model combines a piecewise linear trend with annual and weekly seasonality and accounts for regional public holidays and historical lockdowns. Candidate trend changes can occur throughout the five-year training period, from 1 October 2021 through 30 September 2026. For China's power and residential sectors, the model also accounts for temperature through ERA5 heating and cooling degree days. For the 2026 projections, the heating degree-day record extends through 10 November and the cooling degree-day record through 5 April. We use climatological means for the corresponding day of the year to extend both records through December.

We use the fitted models to project emissions for each day from 1 October to 31 December 2026, with a lower bound of zero. We add these daily projections to the January–September activity estimates to obtain the full-year estimates. We calculate global emissions by summing the 47 series, including global international aviation and shipping once within ROW.

## 2.3 Forecast evaluation and uncertainty

We refit the 47 projection models with the same training period and model specification to estimate forecast uncertainty.

In each of 2,000 simulations, we draw future trend deviations independently for each series and resample centred training residuals in 14-day blocks [38]. Using the same historical dates for all series within each simulation preserves dependence across series and within blocks. We add the simulated deviations to the daily point projections and constrain the resulting emissions to be non-negative. We then sum the fourth-quarter emissions and add the January–September activity estimate. The central 68% of the simulated annual totals gives the forecast interval conditional on the activity estimates and temperature inputs.

September-cutoff hindcasts for 2021–2025 use the same model and simulation procedure, with up to five years of data through 30 September and evaluation over October–December. Training begins in January 2019 for 2021–2023 and uses a full five years for 2024–2025. Hindcast weather forecasts use climatologies calculated only from degree-day data through each cutoff. The seasonal-naive benchmark reuses emissions from the same dates a year earlier. We measure the distance from each realized annual estimate to its simulated interval, taking the distance as zero within the interval. To represent the scale of persistent forecast errors, we widen both ends of the 2026 interval by the largest of these five distances. The adjusted range is an empirical uncertainty summary based on five years.

## *2.4 Annual contributions and daily anomalies*

Annual changes for 2020–2026 are calculated relative to the preceding year for each region-sector series and then aggregated by region and sector. For 2026, we compare the net global change with the sum of all positive and negative regional contributions to quantify the extent of offsetting. Changes are also calculated for four matching periods in 2025 and 2026: January-February, March-April, May-September, and October-December.

We fit separate descriptive baselines to the January 2019–September 2026 activity estimates to capture trends and annual and weekly seasonality in all 47 series. For residential emissions in China, we use a flat trend and a simpler annual seasonal cycle. For Japan, we fit residential emissions on a log scale with a small offset. Daily anomalies during the activity period are departures from the fitted trend and expected annual and weekly seasonality. Climate windows cover June–August and the period following the NOAA/CPC El Niño Advisory on 11 June [12]. Energy windows cover the period from 28 February, the acute March–April disruption, and the period after 17 June, based on the chronology reported by IEA, EIA, and IATA [8, 9, 11]. Comparisons with equivalent windows from 2019–2025 characterize the magnitude, direction, and persistence of daily departures.

Official industrial, vehicle-registration, electricity-demand, and generation statistics support interpretation of the China and India results. Climate and energy-market reports establish the timing and geographic context of the event windows.

## *2.5 Country-level power emissions during heat episodes*

We compare national power emissions in France, Spain, Italy, Germany, the United Kingdom and India during 16–30 June and during independently documented heat dates [39–45]. The French episodes are 17–30 June, 4–19 July and 28 July–19 August. The Spanish episodes are 20–27 June, 2–24 July and 28 July–20 August. The UK uses 16–30 June. Indian bulletins identify regional heat on 11–13 and 17–28 June. We use the common window for Italy and Germany and compare national emissions with the reported regional heat dates in India.

For each window we report cumulative actual-minus-baseline emissions, that departure divided by cumulative baseline emissions, and the change relative to the matching dates in 2025. Baselines use fitted trend and annual and weekly seasonality through 30 September. France uses a positive log-link baseline, back-transformed using the training mean of the exponentiated log-scale residuals. The other primary fits are additive. Sensitivity checks use national-calendar log models fitted through September and log and additive models trained through 31 May. We assess the pre-June designs in 36 held-out country-model-year comparisons in 2023–2025. Weighted absolute percentage error (WAPE) divides the sum of absolute daily errors by total observed emissions over each holdout.

## *2.6 Comparison with other emissions inventories*

We compare complete calendar years from 2019 to 2025 with EDGAR 2026, Global Carbon Project (GCP) fossil $CO_2$ version 2026v11, IEA Global Energy Review 2026 and Climate TRACE v5.12.0 [46–49]. Annual growth uses each source's own preceding-year total as the denominator. National comparisons match reporting areas, calendar years and sector groupings. EU27 is counted as one aggregate with a common 27-member boundary, and the UK is compared separately. Global aggregates include the international aviation and shipping series once.

We map EDGAR fossil $CO_2$ and reported national greenhouse gas inventories (NGHGIs) to the Carbon Monitor IPCC sector definitions [1, 50, 51]. Power covers 1.A.1.a. Industry covers 1.A.1.b–c, 1.A.2 and all category 2 $CO_2$. Ground transport covers road, rail, domestic navigation and other transport (1.A.3.b–e). Residential and other fuel use covers 1.A.4–5. Domestic aviation is 1.A.3.a. Global EDGAR aviation and shipping use the AIR and SEA records (1.C.1–2 in the EDGAR IPCC 1996 workbook). GCP retains coal, oil, gas and cement, with flaring and other process emissions excluded. IEA totals are from the energy-combustion and industrial-process chart, whose scope excludes flaring. IEA growth divides the published increment from the annual-change chart by the preceding year's rounded total [52]. The NGHGI comparison sums the five domestic sectors for EU27, Brazil, Russia, the United Kingdom and the United States, whose matched records cover every year from 2019 to 2022.

Using the Climate TRACE IPCC crosswalk [53], we include generation, heating plants and remaining 1.A.1.a energy use in power. Ground transport includes road, rail, domestic navigation and other transport. The buildings and other fuel-use sector includes residential, non-residential and unspecified onsite fuel use. Industry includes manufacturing, non-fuel mining, quarrying and the reported refinery aggregate. This aggregate, assigned to 1.A.1.b in the crosswalk, also contains process, fugitive and flaring $CO_2$. Mixed upstream oil-and-gas emissions lack separable 1.A.1.c and 1.B.2 components in the aggregate. We exclude this mixed series, solid-fuel fugitive sources and vessels with unallocated domestic or international status. Domestic aviation and the two international series use the corresponding Climate TRACE categories. We sum twelve months of direct $CO_2$ and exclude Scope 2 estimates.

Figure 1d–g shows the comparison for 11 countries and EU27. EDGAR and Climate TRACE use the five mapped domestic sectors. The IEA Greenhouse Gas Emissions from Energy Data Explorer provides 2024 fuel-combustion estimates for power and heat, industry, domestic transport and buildings [54]. For IEA domestic transport, we combine Carbon Monitor ground

transport and domestic aviation. GCP country totals from coal, oil, gas and cement are compared with the sum of the five domestic Carbon Monitor sectors. Percentage differences use Carbon Monitor as the denominator.

# 3. Results

## *3.1 Comparisons with other emissions inventories*

All five global series decline in 2020, rebound in 2021 and rise in each subsequent year through 2025 (Fig. 1a,b). In 2025, Carbon Monitor estimates 38.0 Gt $CO_2$ (+0.6%), compared with 38.1 Gt in GCP (+0.9%), 38.5 Gt in mapped EDGAR (+0.4%), 38.1 Gt in IEA (+0.4%) and 39.4 Gt in the seven mapped Climate TRACE sectors (+0.4%). Carbon Monitor exceeds reported NGHGI emissions by 32.6–34.8 Mt $CO_2$ per year across the same five reporting areas and five domestic sectors in 2019–2022 (Fig. 1c).

Differences vary across countries and sectors (Fig. 1d–g). For example, in 2025 EDGAR estimates ground transport emissions 9.32% below Carbon Monitor in China and 6.28% below in India. Climate TRACE estimates are 13.5% higher in China and 47.6% lower in India. In the 2024 IEA comparison, the industry series covers fuel combustion, while the mapped Carbon Monitor industry series also includes process $CO_2$.

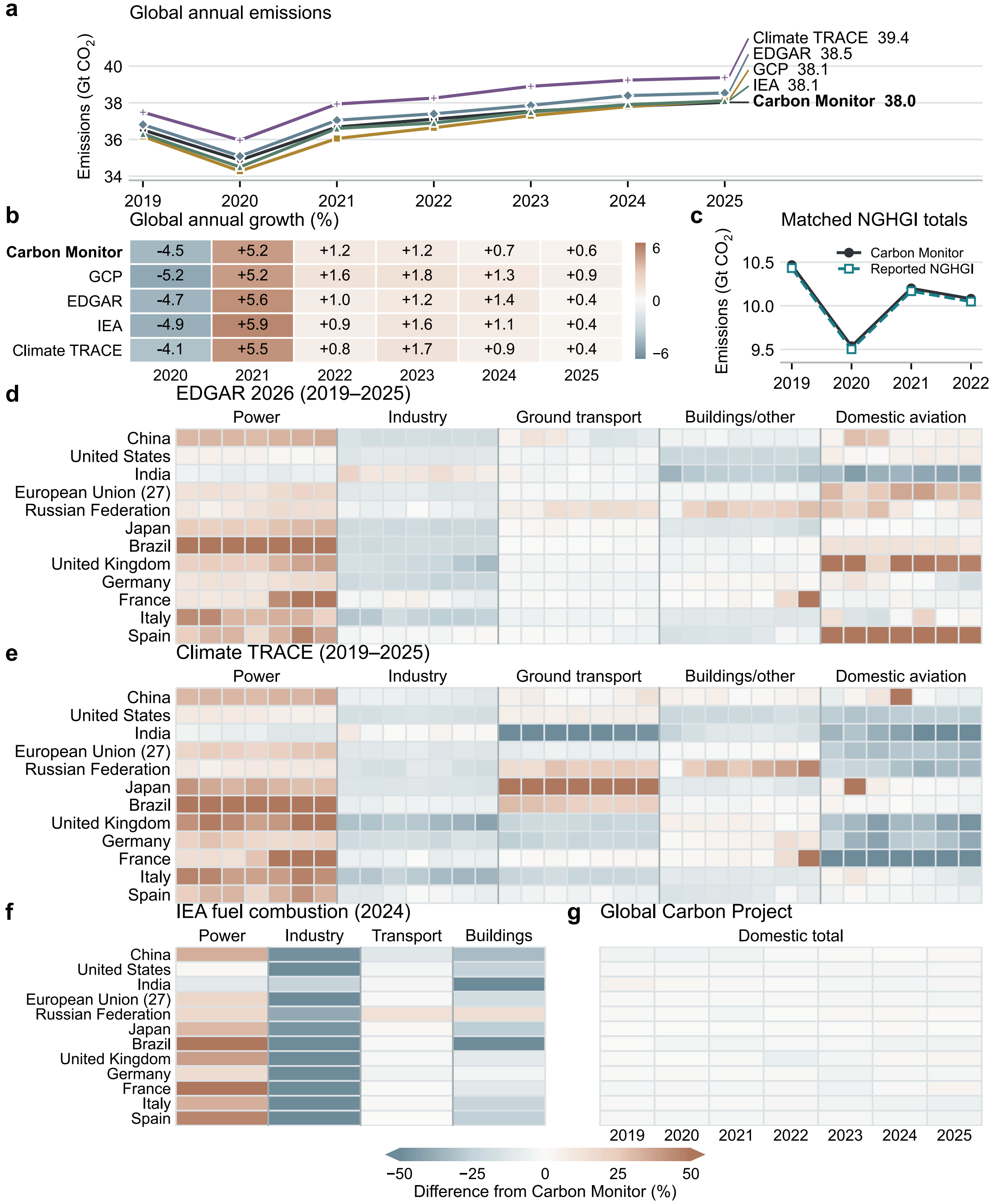

*Figure 1 | $CO_2$ emissions across Carbon Monitor and other inventories. a, Global annual emissions in 2019–2025. Labels give 2025 totals in Gt $CO_2$. b, Global year-on-year growth in each annual series. c, Combined $CO_2$ from five domestic sectors for EU27, Brazil, Russia, the United Kingdom and the United States in Carbon Monitor and the reported national greenhouse gas inventories (NGHGI). d–g, Annual differences relative to Carbon Monitor for 11 countries and EU27. EDGAR (d) and Climate TRACE (e) use the five mapped domestic sectors for 2019–2025. IEA (f) provides fuel-combustion data for 2024. GCP (g) reports domestic totals for 2019–2025. In panels d–g, blue indicates lower estimates than Carbon Monitor and brown indicates higher estimates. The color scale is capped at ±50%. Section 2.6 defines the comparison boundaries.*

## *3.2 Global annual emissions remain close to the 2025 level*

Global emissions are estimated at 38.0 Gt $CO_2$ in 2026, a change of −34.1 Mt (−0.1%) from 2025 (Fig. 2 and Table 1). The 2026 estimate corresponds to 104 Mt $CO_2$ d-1. Activity-based estimates through 30 September sum to 28.2 Gt, 23.0 Mt less than in the same months of 2025. The October–December projection is 9.74 Gt, 11.1 Mt less than in the same months of 2025.

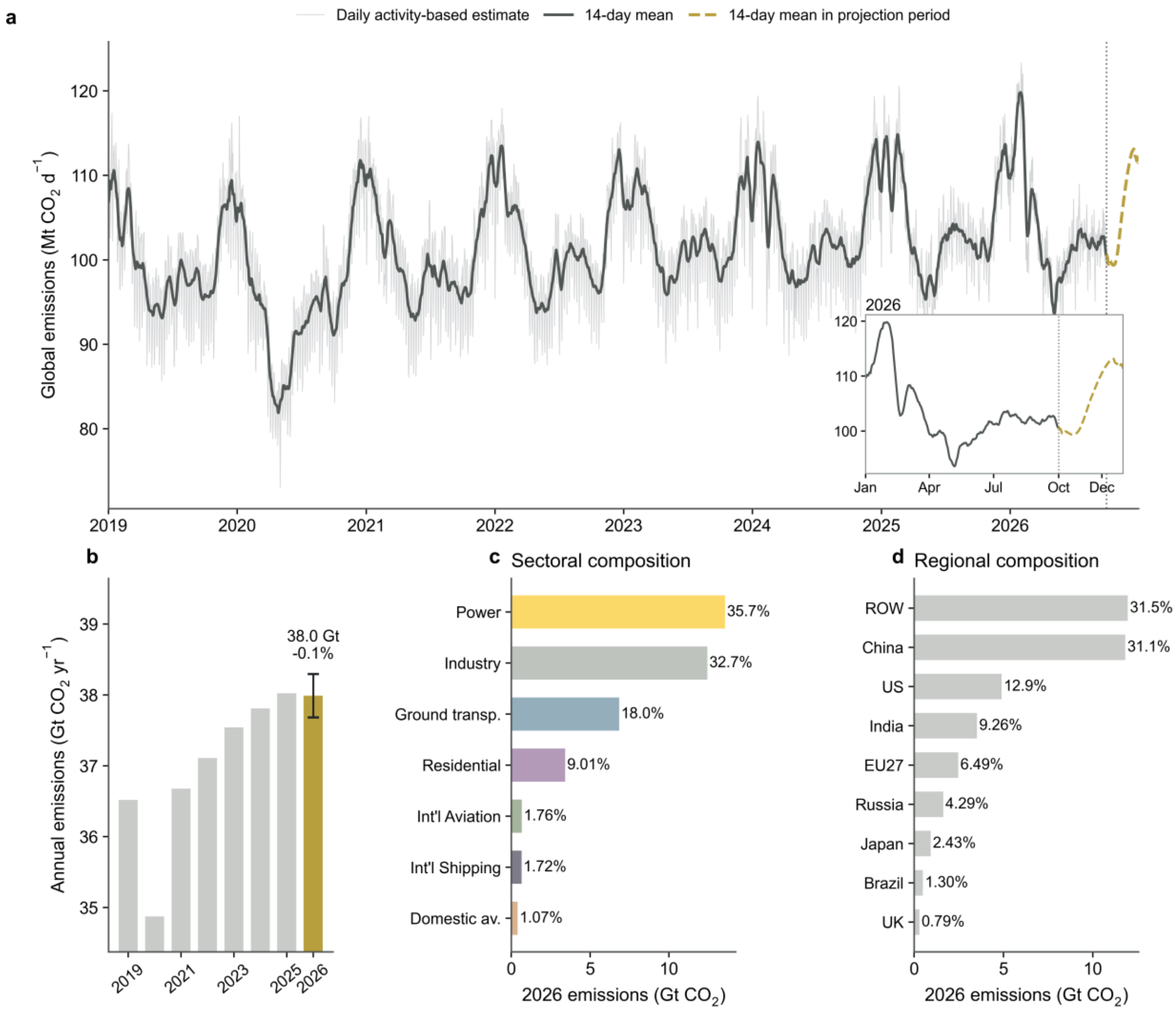

*Figure 2 | Global daily and annual $CO_2$ emissions, 2019–2026. a, Daily activity-based estimates through 30 September 2026 and a centred 14-day moving mean of the full series, dashed for the projection period. Windows spanning 1 October include both activity estimates and projections. The inset shows 2026, and the dotted line marks 1 October. b, Annual totals. The whisker shows the hindcast-adjusted forecast range for 2026. c, Sector shares of the 2026 estimate. d, Regional shares.*

***Table 1 | Global annual $CO_2$ emissions, 2019–2026.*** *Growth rates use unrounded totals. The 2026 row combines activity estimates through September with Prophet projections for the final 92 days. Its hindcast-adjusted forecast range is 37.7–38.3 Gt $CO_2$.*

| Year | Total (Gt $CO_2$) | Daily mean (Mt d-1) | Year-on-year (%) | Basis |
|---|---|---|---|---|
| 2019 | 36.5 | $1.00\times10^{2}$ | NA | Activity estimate |
| 2020 | 34.9 | 95.3 | -4.5 | Activity estimate |
| 2021 | 36.7 | $1.00\times10^{2}$ | +5.2 | Activity estimate |
| 2022 | 37.1 | 102 | +1.2 | Activity estimate |
| 2023 | 37.5 | 103 | +1.2 | Activity estimate |
| 2024 | 37.8 | 103 | +0.7 | Activity estimate |
| 2025 | 38.0 | 104 | +0.6 | Activity estimate |
| 2026 | 38.0 | 104 | -0.1 | Activity and projection |

Regional movements are much larger than the global net change. Increasing regions add 0.331 Gt and declining regions subtract 0.365 Gt, cancelling 95.1% of their combined 0.696 Gt movement. Relative to the same 2025 periods, global emissions change by +52.7 Mt in January–February, +4.09 Mt in March–April, −79.8 Mt in May–September, and a projected −11.1 Mt in October–December.

Replacing the October–December projection with emissions from the same dates in 2025 gives a year-on-year change of −0.1%. The seasonal-naive estimate is 11.1 Mt higher than the Prophet estimate.

## 3.3 China's reductions are concentrated in industry and ground transport

China emits an estimated 11.8 Gt in 2026 (31.1% of the global total). Its annual change of −243 Mt (−2.0%) is the largest regional reduction (Table 2 and Fig. 3). China accounts for 66.6% of gross regional reductions.

***Table 2 | Regions and year-on-year changes in 2026.*** *Global international aviation and shipping are assigned to ROW once.*

| Region | 2026 total (Gt) | Global share (%) | Change (Mt) | Change (%) |
|---|---|---|---|---|
| China | 11.8 | 31.1 | -243 | -2.0 |
| United States | 4.90 | 12.9 | -54.6 | -1.1 |
| India | 3.52 | 9.26 | +222 | +6.7 |
| EU27 | 2.46 | 6.49 | +36.8 | +1.5 |
| Russian Federation | 1.63 | 4.29 | -36.4 | -2.2 |
| Japan | 0.92 | 2.43 | -24.8 | -2.6 |
| Brazil | 0.49 | 1.30 | +0.37 | +0.1 |
| United Kingdom | 0.30 | 0.79 | -6.22 | -2.0 |
| ROW | 12.0 | 31.5 | +72.1 | +0.6 |

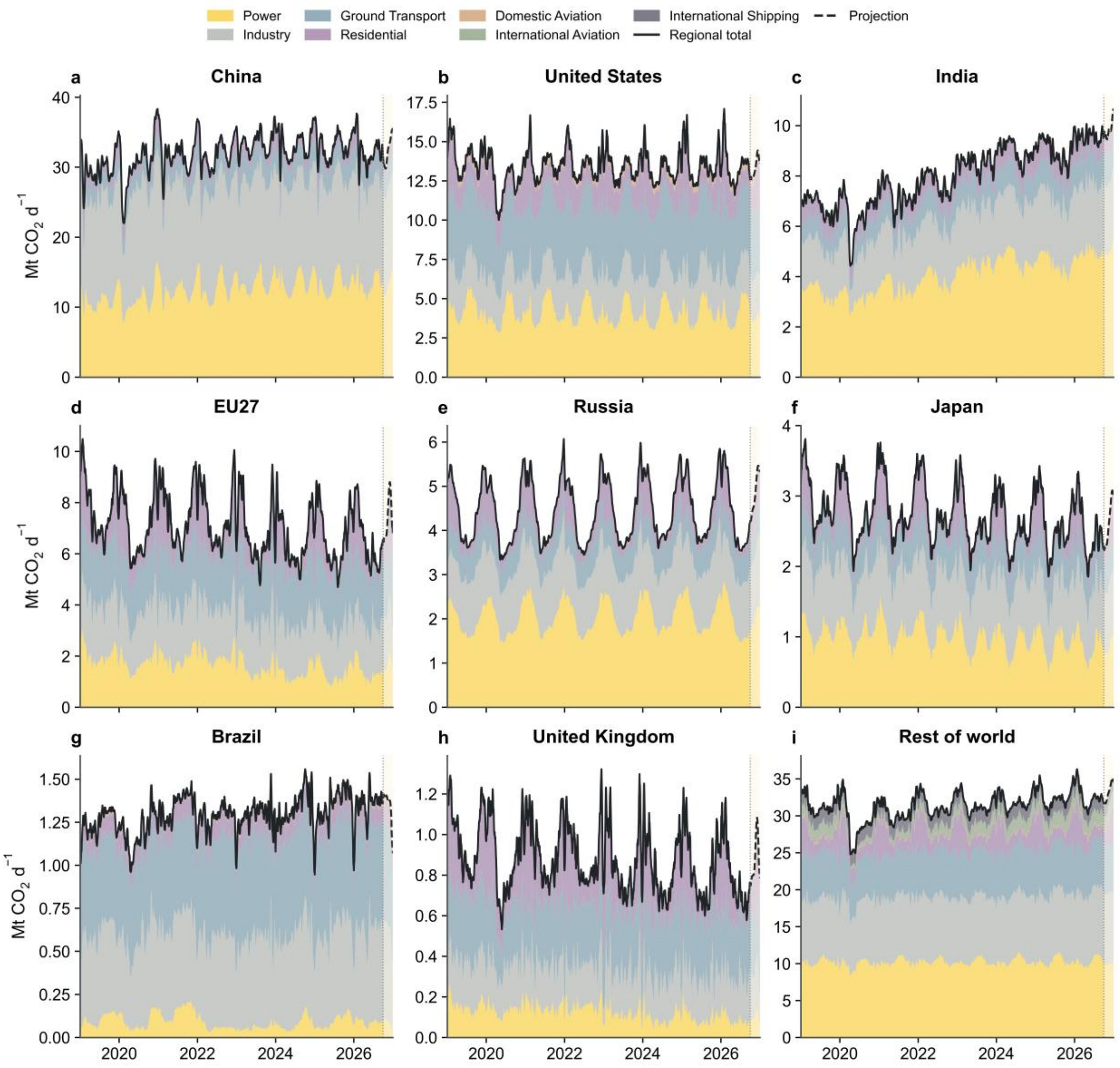

*Figure 3 | Regional daily trajectories and sectoral composition. Colored areas show centred 14-day moving means of sectoral emissions, and the outline shows their regional total. Windows spanning 1 October include both activity estimates and projections. Lighter shading and dashed totals identify the projection period beginning on 1 October 2026. The regional axes have independent scales. Global international aviation and shipping enter once through ROW.*

China's largest reductions are in industry (−157 Mt) and ground transport (−44.9 Mt), followed by residential fuel use (−24.9 Mt) and power (−16.8 Mt). Domestic aviation increases by 0.967 Mt (Fig. 4).

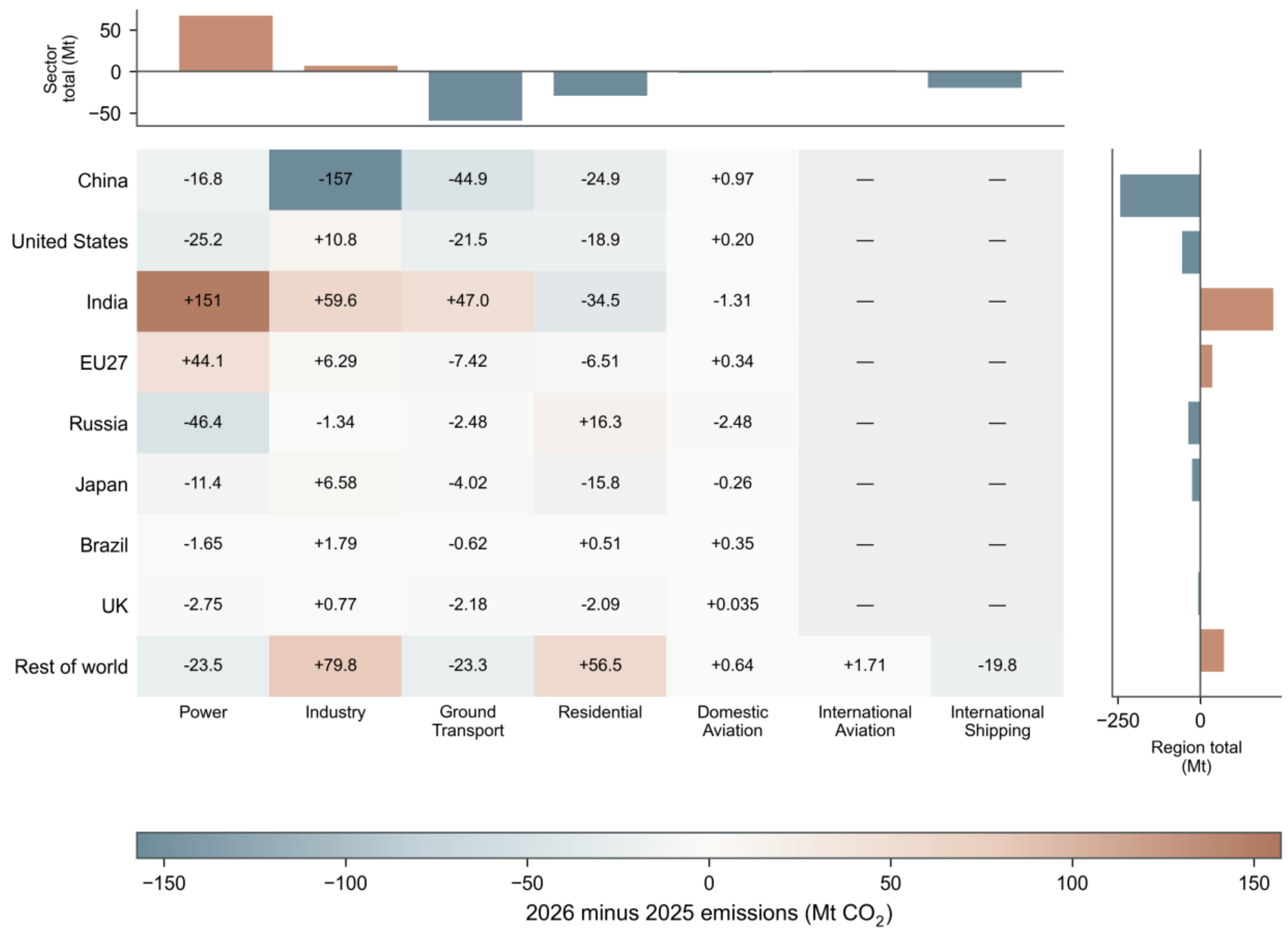

*Figure 4 | Region-sector contributions to the 2026 annual change. Cells show the 2026 minus 2025 change for each of the 47 region-sector series. Global international aviation and shipping are counted once and assigned to ROW. Marginal bars summarize regional and sectoral totals. China's industrial and ground-transport reductions and India's power and industrial increases dominate the opposing movements.*

China's industrial emissions change by +12.5 Mt in January–February, −30.4 Mt in March–April, −82.9 Mt in May–September, and a projected −56.6 Mt in October–December. Ground transport changes by +4.45 Mt, +4.31 Mt, −48.4 Mt and a projected −5.31 Mt over the same periods.

Official Chinese activity indicators provide context for these divergent sector estimates. Value added in above-designated-size industrial enterprises and their manufacturing sector grew during January-August, while crude-oil processing, motor-vehicle output, cement, and rolled-steel output declined in the relevant official releases. New-energy vehicles accounted for 49.4% of new automobile registrations in the first half of 2026, although they were 13.2% of the automobile fleet at the end of June [55, 56]. The emissions pattern is consistent with lower activity in several energy-intensive industries alongside continued growth in manufacturing value added. Fleet electrification provides additional context for the ground-transport decline.

## *3.4 India's power and industrial growth offset much of China's decline*

India records the largest regional increase, +222 Mt (+6.7%), reaching 3.52 Gt. It accounts for 67.0% of gross regional increases. The largest sectoral increases are power (+151 Mt), industry (+59.6 Mt) and ground transport (+47.0 Mt).

India's power emissions change by −0.87 Mt in January–February, +0.56 Mt in March–April, +82.4 Mt in May–September, and a projected +68.9 Mt in October–December. Industrial emissions change by +6.60 Mt, +6.21 Mt, +27.7 Mt and a projected +19.1 Mt across these periods. Official data report strong electricity-consumption growth over 1–27 April, a then-record peak demand of 256 GW on 25 April, and year-on-year increases in industrial production, manufacturing, and electricity and gas supply in August. Coal and lignite accounted for 69.5% of total electricity supplied during April–June, although non-fossil capacity exceeded 54% of installed capacity by the end of July [57–60].

ROW contributes +72.1 Mt (+0.6%). Industry (+79.8 Mt) and residential fuel use (+56.5 Mt) account for most of the increase, while power (−23.5 Mt) and ground transport (−23.3 Mt) provide the largest reductions. The ROW accounting group includes global international aviation and shipping.

EU27 gains 36.8 Mt (+1.5%) and Brazil is nearly unchanged at +0.375 Mt (+0.1%). Declines are projected for the United States (−54.6 Mt, −1.1%), Russia (−36.4 Mt, −2.2%), Japan (−24.8 Mt, −2.6%) and the United Kingdom (−6.22 Mt, −2.0%).

## *3.5 Sectoral balance is dominated by lower ground transport*

The four largest sectors in 2026 are power at 13.6 Gt (35.7%), industry at 12.4 Gt (32.7%), ground transport at 6.85 Gt (18.0%) and residential fuel use at 3.42 Gt (9.01%) (Table 3).

***Table 3 | Sectors and year-on-year changes in 2026.*** *Residential covers IPCC categories 1.A.4–5.*

| Sector | 2026 total (Gt) | Global share (%) | Change (Mt) | Change (%) |
|---|---|---|---|---|
| Power | 13.6 | 35.7 | +67.3 | +0.5 |
| Industry | 12.4 | 32.7 | +6.83 | +0.1 |
| Ground Transport | 6.85 | 18.0 | -59.4 | -0.9 |
| Residential | 3.42 | 9.01 | -29.3 | -0.8 |
| Domestic Aviation | 0.40 | 1.07 | -1.53 | -0.4 |
| International Aviation | 0.67 | 1.76 | +1.71 | +0.3 |
| International Shipping | 0.65 | 1.72 | -19.8 | -2.9 |

Ground transport supplies the largest sectoral reduction, −59.4 Mt (−0.9%), followed by residential fuel use (−29.3 Mt) and international shipping (−19.8 Mt). Power contributes the largest increase at +67.3 Mt (+0.5%). Industry and international aviation increase by 6.83 Mt and 1.71 Mt, respectively.

The region-sector matrix shows how these changes combine across regions (Fig. 4). In power, India contributes +151 Mt and the United States contributes −25.2 Mt.

The annual record puts the 2026 changes in context (Table 4). Global ground transport emissions had increased in every year from 2021 to 2025 before the projected 2026 decline. Global power emissions rebound by 67.3 Mt after falling by 142 Mt in 2025. China’s estimated decline in 2026 follows reductions of 47.3 Mt in 2024 and 99.4 Mt in 2025, while India’s increase rises from 12.6 Mt in 2025 to 222 Mt in 2026.

***Table 4 | Annual changes in $CO_2$ emissions by region and sector, 2020–2026 (Mt $CO_2$).*** *Values are differences from the preceding calendar year. Regional and sectoral rows are alternative decompositions of the global change. ROW includes global international aviation and shipping. The 2026 column includes October–December projections.*

| **Country, region or sector** | **2020** | **2021** | **2022** | **2023** | **2024** | **2025** | **2026** |
|---|---|---|---|---|---|---|---|
| **Country/region** | | | | | | | |
| China | +224 | +426 | +91.1 | +488 | -47.3 | -99.4 | -243 |
| United States | -527 | +334 | +36.3 | -105 | -17.0 | +92.3 | -54.6 |
| India | -109 | +259 | +192 | +284 | +167 | +12.6 | +222 |
| EU27 | -276 | +176 | -64.7 | -221 | -74.9 | +15.6 | +36.8 |
| Russian Federation | -69.7 | +80.7 | -54.4 | +22.5 | +49.9 | -2.15 | -36.4 |
| Japan | -64.1 | +20.9 | -26.3 | -52.7 | -19.4 | -1.99 | -24.8 |
| Brazil | -18.4 | +53.2 | -23.4 | +4.64 | +15.1 | +6.47 | +0.37 |
| United Kingdom | -38.4 | +18.7 | -12.7 | -16.1 | -6.56 | -0.36 | -6.22 |
| Rest of world | -764 | +434 | +295 | +27.2 | +201 | +191 | +72.1 |
| **Global total** | $-1.64\times10^{3}$ | $+1.80\times10^{3}$ | +432 | +431 | +268 | +214 | -34.1 |
| **Sector** | | | | | | | |
| Power | -394 | +965 | +193 | +129 | +201 | -142 | +67.3 |
| Industry | +68.0 | +254 | -166 | +137 | -36.8 | +212 | +6.83 |
| Ground Transport | -668 | +466 | +274 | +141 | +41.2 | +24.5 | -59.4 |
| Residential | -137 | +13.9 | -27.1 | $-1.60\times10^{2}$ | -4.10 | +77.5 | -29.3 |
| Domestic Aviation | -125 | +53.0 | +11.9 | +59.3 | +15.5 | +8.64 | -1.53 |
| International Aviation | -329 | +30.7 | +145 | +99.3 | +63.4 | +32.1 | +1.71 |
| International Shipping | -57.3 | +21.6 | +1.99 | +25.4 | -11.7 | +1.10 | -19.8 |

## 3.6 Regional contrasts during the climate and energy windows

The common screening procedure identifies both positive and negative anomalies across the 47 series (Fig. 5). During June–August, departures from the fitted baseline range from −10.7% to +8.80%. Global power is +0.21% relative to its fitted baseline over June–August. EU27 power departs by +2.47% in June and −4.45% in July.

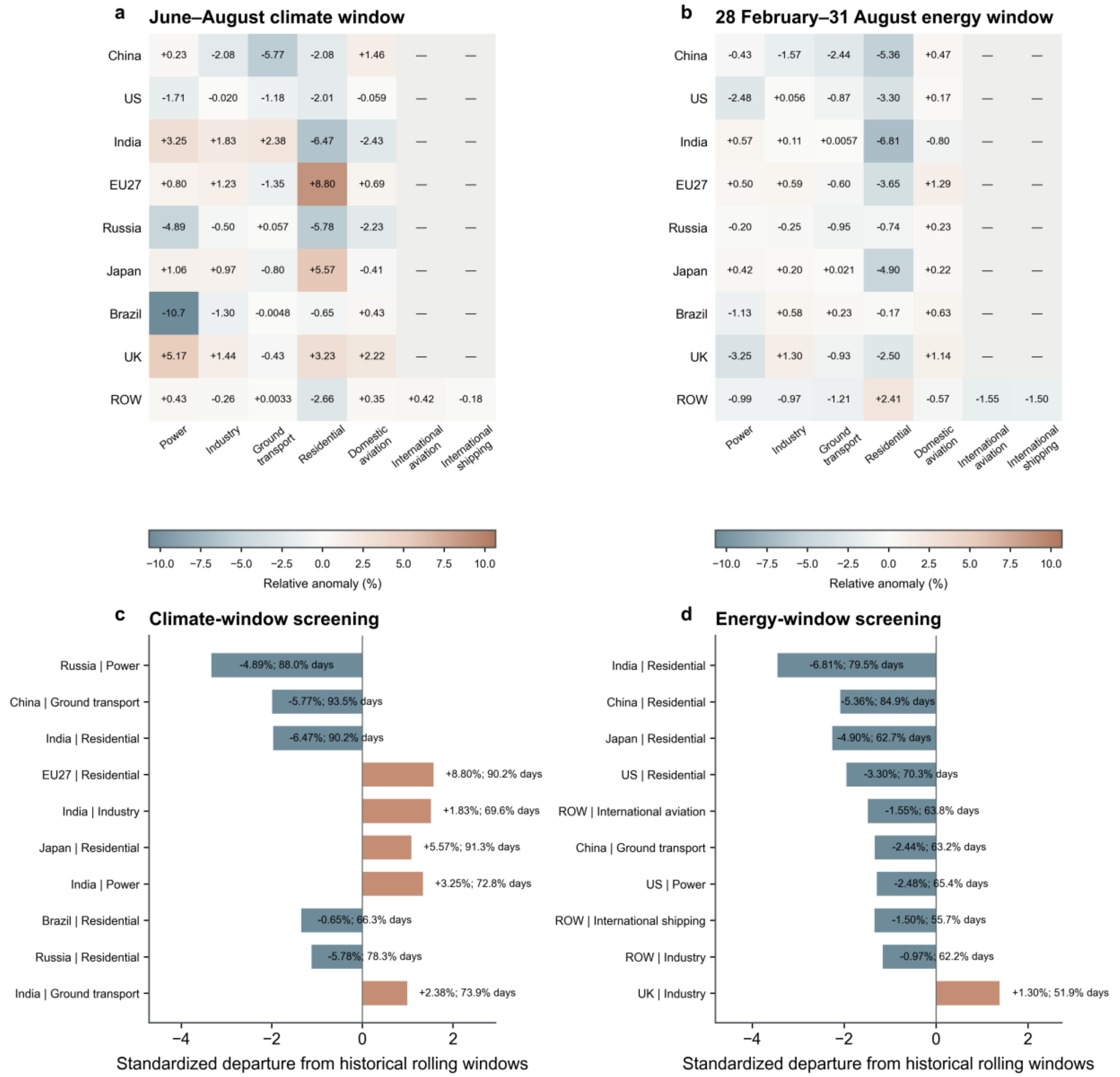

*Figure 5 | Regional and sectoral anomalies during the 2026 climate and energy windows. a,b, Relative anomalies from fitted trend and expected annual and weekly seasonality for all 47 series during June-August and 28 February-31 August. c,d, The ten highest screening scores in each window. Scores combine the absolute standardized departure from rolling 2019-2025 windows of equal length with the share of days having the dominant sign. Bar labels give the relative anomaly and the percentage of days with that sign.*

These emissions contrasts occurred during exceptional summer warmth reported by Copernicus in western Europe and several subregions of Asia, North America and Africa [13, 14]. The screening describes how daily emissions moved relative to their fitted baselines within the selected windows.

Country-level power diagnostics resolve changes muted in the EU27 and seasonal aggregates (Figs. 6 and 7). During the common 16-30 June window, departures are +124% in France, +47.8% in the United Kingdom, +17.3% in Italy, +7.51% in India, +4.01% in Germany and +5.03% in Spain; the EU27 reference is +15.0%. France has the largest relative departure among these countries, but India's cumulative departure (5.51 Mt) exceeds France's (0.32 Mt), reflecting their different emission baselines. The common dates compare national series and do not impose identical national heatwave boundaries.

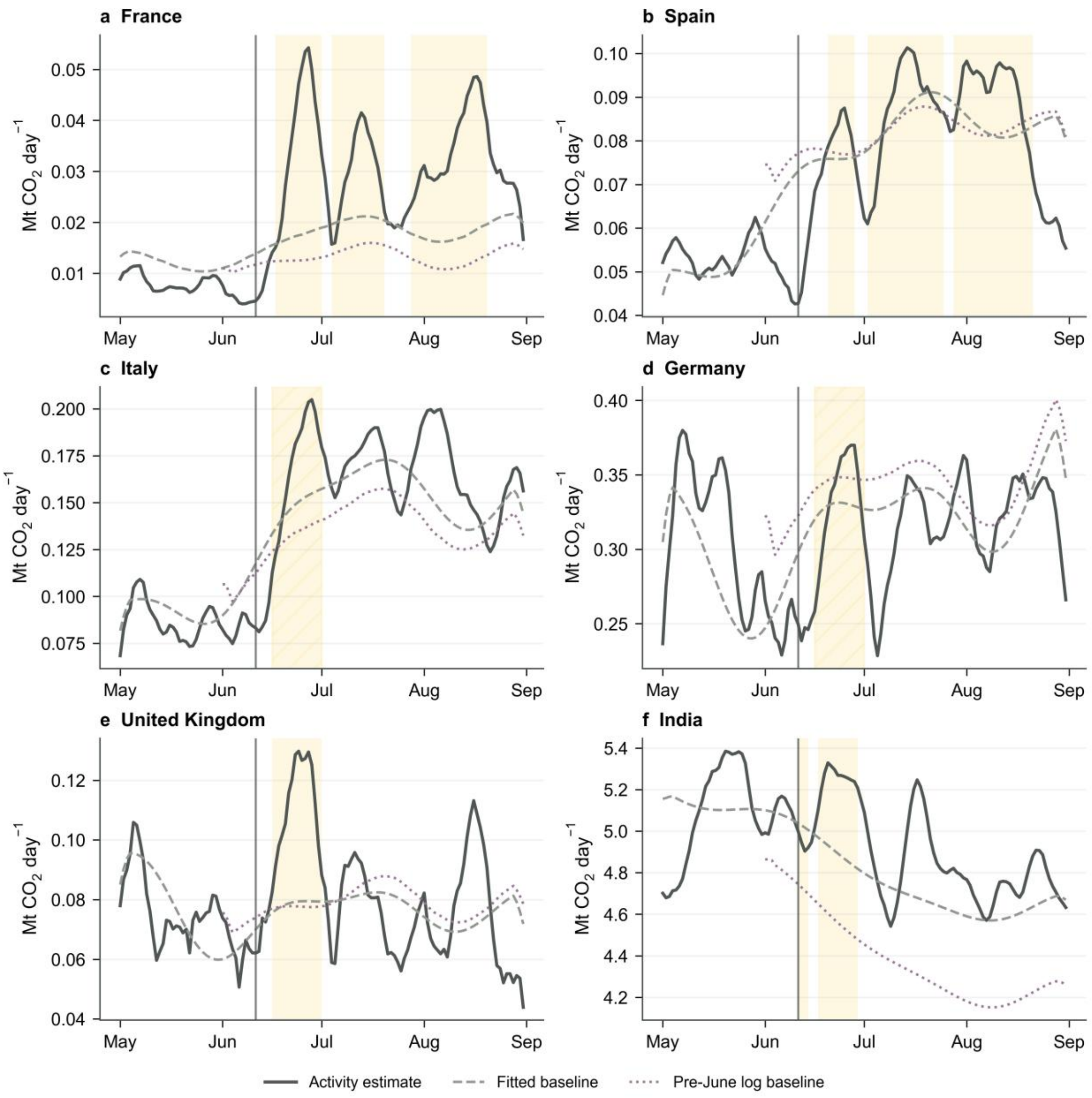

*Figure 6 | Country-level power emissions and heat-period timing in 2026. a, France; b, Spain; c, Italy; d, Germany; e, United Kingdom; f, India. Curves show centered seven-day means on independent vertical scales. Charcoal curves are activity estimates; grey dashed curves are baselines fitted through 30 September. The French fitted baseline uses a positive log link; the other grey baselines are additive. Dusty-purple dotted curves use national-calendar log models trained through 31 May and are shown only for June-August. Yellow shading marks documented June-August heat dates; hatched shading for Italy and Germany marks the common 16-30 June comparison window, whose dates are not country-specific national heatwave boundaries. India's shading concerns regional heat records compared with national emissions. The vertical line marks the 11 June NOAA/CPC advisory, not an exact physical ENSO onset or attribution boundary. Reported departures use unsmoothed daily sums; no causal effect or uncertainty band is implied by the plotted curves.*

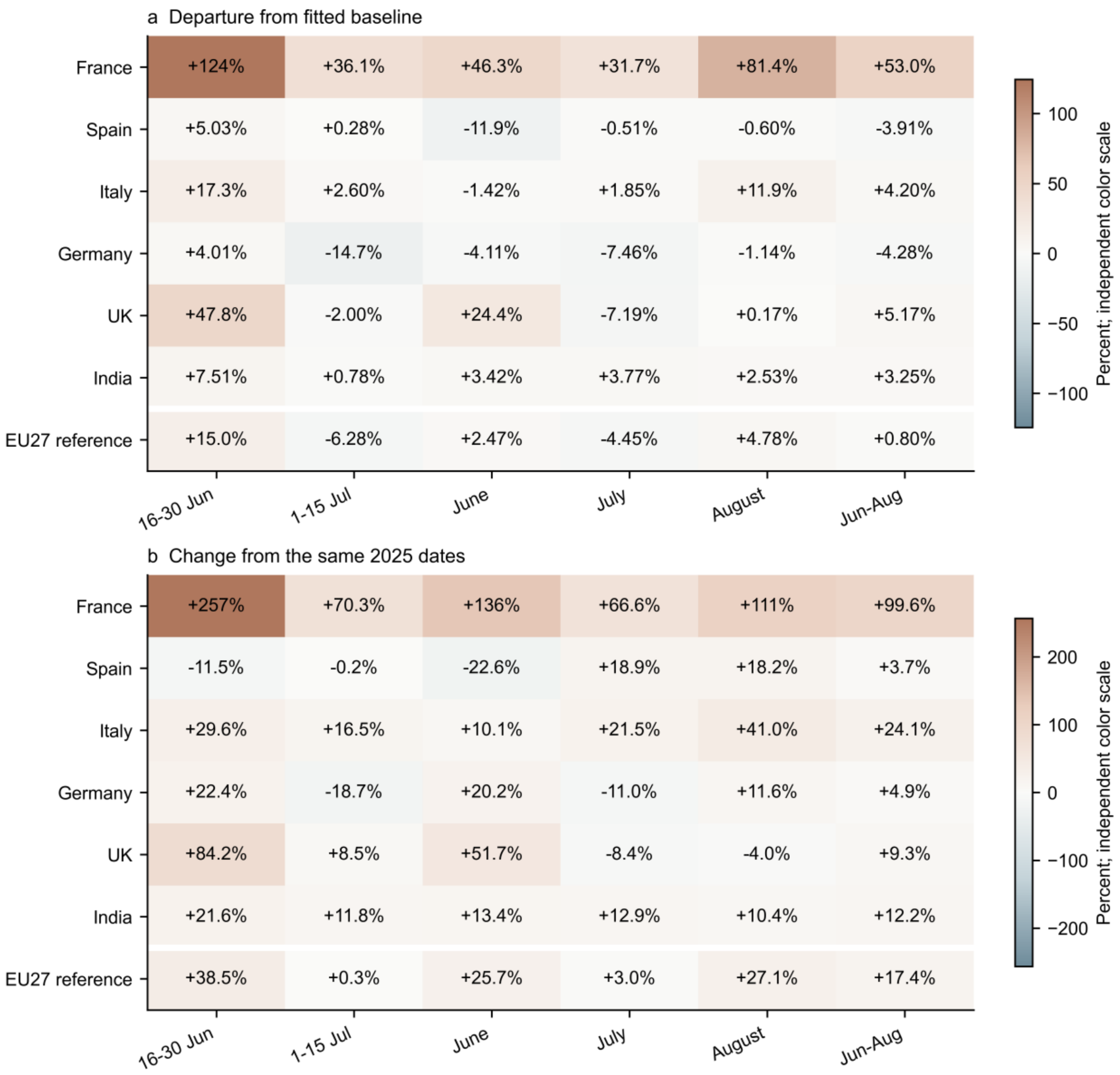

*Figure 7 | Country-level power departures across short and seasonal comparison windows in 2026. a, Cumulative activity estimates relative to their trend-plus-seasonality baselines. b, Changes from the same calendar dates in 2025. The panels have independent color scales. The common windows do not assert uniform national heat exposure, and India's row is a same-date comparison. France uses a positive-link baseline and the other primary country baselines are additive. EU27 is a reference aggregate overlapping four displayed countries and is not added to them. Monthly and summer averages can mask shorter positive departures; results describe temporal associations.*

Using independently documented heat dates changes the seasonal interpretation. France's three summer heatwaves span 53 unique days with a +95.6% departure (+0.92 Mt). Spain's 55-day heatwave union has a +8.32% departure, compared with -3.91% over June-August. The UK's 16-30 June heat episode has a +47.8% departure, compared with +5.17% over the summer.

During 17-28 June, when successive India Meteorological Department (IMD) reports document regional heat on every date, India's national power emissions are +8.25% above baseline (+4.85 Mt) and +22.8% above matching 2025 dates. The observed-day Indian union also includes 11-13 June, but local exposure and national emissions remain different spatial scopes. Thus, neither the summer mean nor year-on-year growth is interchangeable with the event-specific baseline departure.

Positive 16–30 June departures persist across valid fits in France (+124% to +207%), the UK (+47.2% to +56.0%), Italy (+17.3% to +30.3%) and India (+7.10% to +15.3%). The departure falls to −0.89% in Germany under the pre-June log fit and to −0.50% in Spain under the pre-June additive fit. Mean WAPE in the 2023–2025 summer holdouts is 6.32–33.4% across the national log models. In France, the additive model's mean WAPE is 113%, compared with 33.4% for the log model, and the additive baseline is nonpositive on 118 held-out days.

The energy-disruption window from 28 February through 31 August spans anomalies from −6.81% to +2.41%. Global ground transport averages −1.08% relative to baseline over this window. International aviation and shipping have anomalies of −4.99% and −5.01% during March–April, and +1.12% and −3.76% in August. IATA reports lower April passenger demand concentrated in the Middle East. IEA and EIA document disruptions to energy and shipping markets [8–11].

## 3.7 Forecast evaluation and uncertainty in the annual estimate

Five September-cutoff hindcasts for 2021–2025 evaluate the projection design (Fig. 8). Mean daily MAE and RMSE are 2.63 and 3.30 Mt $CO_2$ d-1 for Prophet and 3.69 and 4.58 Mt $CO_2$ d-1 for the seasonal-naive benchmark. Mean absolute annual biases are 154 and 66.7 Mt, respectively (Table 5).

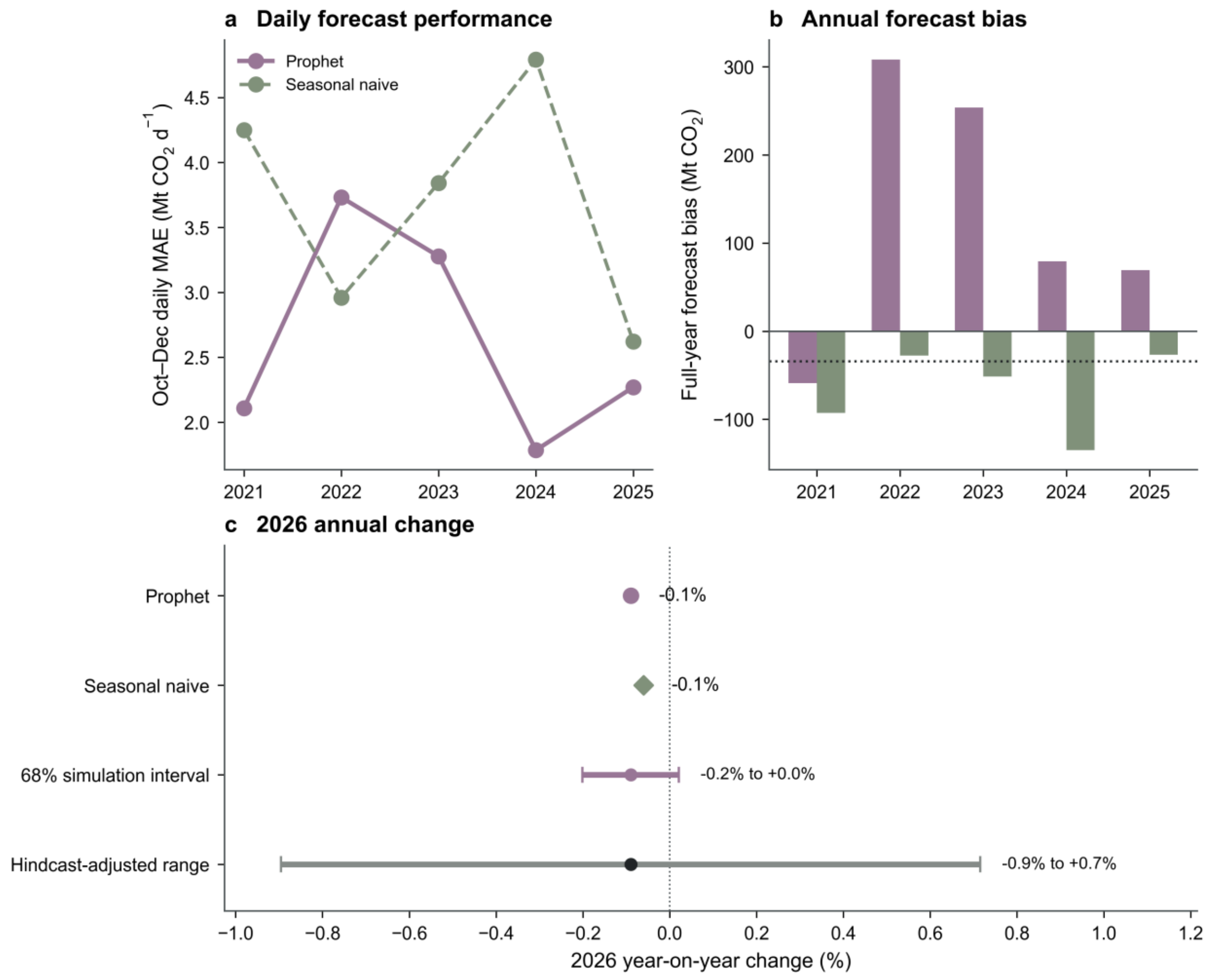

*Figure 8 | Forecast evaluation and uncertainty in the 2026 annual estimate. a, October–December daily mean absolute error for Prophet and the same-date seasonal-naive benchmark in five September-cutoff hindcasts. b, Full-year forecast biases in those years. The dotted line marks the Prophet point estimate of the 2026 annual change. c, Prophet and seasonal-naive point estimates for 2026. The coloured interval gives the central 68% of the forecast simulations. The grey range widens those bounds by the largest amount by which a realized annual estimate lay outside its simulated interval in the five hindcasts.*

***Table 5 | September-cutoff hindcast performance, 2021–2025.*** *Metrics are averaged across the five hindcasts. Each evaluates 92 projected days. APE denotes absolute percentage error. Interval hits count the years in which the realized annual Carbon Monitor estimate falls within the model-based interval.*

| Model | Daily MAE (Mt) | Daily RMSE (Mt) | Mean absolute annual bias (Mt) | Mean annual APE (%) | 68% interval hits |
|---|---|---|---|---|---|
| Prophet | 2.63 | 3.30 | 154 | 0.41 | 0/5 |
| Seasonal-naive | 3.69 | 4.58 | 66.7 | 0.18 | – |

The central 68% conditional simulation interval for 2026 is 37.9–38.0 Gt $CO_2$, corresponding to −0.2% to +0.0% year on year. The upper bound is slightly above zero before rounding. The simulated interval covers the realized annual Carbon Monitor estimate in 0 of five hindcasts. The largest distance from a realized annual estimate to its simulated interval is 264 Mt. Widening the 2026 bounds by that amount gives 37.7–38.3 Gt $CO_2$, or −0.9% to +0.7% year on year.

# 4. Discussion

## 4.1 Global stabilization and the breadth of mitigation

A global emissions plateau can emerge while economies remain at different stages of transition. Reductions concentrated in particular regions or sectors can bring annual growth close to zero without a simultaneous leveling of emissions elsewhere. The contribution analysis makes this distinction visible. It provides a basis for assessing the breadth of mitigation progress alongside the global growth rate, which alone can conceal both substantial reductions and continuing expansion.

The within-year decomposition shows that the global decline was concentrated in May–September after increases earlier in the year. We project a smaller additional decline in the fourth quarter. Following these components as activity estimates become available will help determine whether the plateau is becoming more broadly established.

## 4.2 Different balances between activity growth and decarbonization

China's sectoral pattern illustrates why industrial value added and emissions need to be considered together with physical output. Growth in manufacturing value added can coexist with declining production in energy-intensive activities [55]. The resulting emissions trajectory may therefore differ from the trajectory of aggregate economic activity. In transport, changes in fleet composition provide a complementary perspective on fuel demand: rapid growth in new-energy-vehicle registrations is relevant to interpreting the observed ground-transport decline [56]. The persistence of these reductions will depend on the balance between changing technology, the mix of production, and activity growth [61]. This balance matters for provincial mitigation pathways and China's national carbon budget [62, 63].

India's results concern the rate at which low-carbon electricity supply can meet rising demand. Industrial expansion and high electricity demand increase the amount of generation required [57, 58]. When coal and lignite retain a large generation share, additional non-fossil capacity can coexist with higher fossil generation [59, 60]. Generation delivered, rather than installed capacity alone, therefore determines how quickly new supply translates into lower power-sector emissions.

Together, these contrasts show why the timing of regional transitions matters for the global trajectory. Sustained global reductions will require declining emissions in established sources to be accompanied by lower-carbon ways of meeting demand in expanding economies.

### *4.3 Interpreting climate and energy-window contrasts*

Annual totals can obscure the duration and recovery of short-lived disruptions. The aviation and shipping profiles distinguish an acute interruption from the subsequent return of activity, while IATA and IEA/EIA records provide the corresponding transport and energy-market context [8–11]. Examining the full sequence helps assess whether an event-window deficit remains visible in the annual balance.

Climate interpretation benefits from resolving both national differences and event timing [64]. The Spanish heat-day union has a positive power departure despite a negative summer average, while the UK's strong June departure weakens in its seasonal mean. French percentages are large on a low power-emissions baseline, whereas India's larger absolute departures occur against much higher national emissions. National emissions also cover areas outside the locally reported Indian heat exposure. NOAA's ENSO classification and Copernicus temperature reports establish the large-scale background [12–14], but do not identify an ENSO contribution to individual heat episodes. The country diagnostics therefore identify periods for joint analysis of temperatures, cooling demand, generation by fuel, plant availability, hydrology and cross-border electricity exchange [65].

### 4.4 What forecast evaluation contributes to the annual assessment

Annual bias accumulates signed daily errors, so a persistent directional offset can affect the annual sum even when typical daily deviations are relatively small. In the hindcasts, Prophet has smaller mean daily errors and larger annual biases than the seasonal-naive benchmark. The daily path helps resolve the unfinished year's temporal structure, and accumulated errors inform interpretation of the annual total.

The Prophet projection and the seasonal-naive benchmark both place global emissions close to the 2025 level. The simulated trend changes and short-run residuals yield too little dispersion to cover the observed annual errors in these hindcasts. The hindcast-adjusted range in Fig. 8 includes both directions of change and its width depends on the five historical hindcasts.

### 4.5 Scope and interpretation

The study covers fossil-fuel combustion and industrial-process $CO_2$ represented in Carbon Monitor. Event-window comparisons describe temporal associations. Residential estimates incorporate temperature-sensitive activity information, so weather and fuel-use records would help distinguish mechanisms within regional changes.

The hindcasts use a consistent retrospective activity release. Their errors describe the historical performance of the forecasting specification on that release. Uncertainty in annual changes also depends on how inventory estimation errors persist between years.

The historical inventory comparison shows similar global annual trajectories through 2025 alongside differences in levels for individual countries and sectors. Sector boundaries and estimation methods can affect agreement [66]. National greenhouse gas inventories serve as historical anchors for Carbon Monitor [1]. The September 2026 Carbon Monitor release also uses EDGAR 2025 as a baseline for parts of the rest-of-world series [67]. Agreement with either inventory therefore partly reflects shared inputs.

# 5. Conclusion

Global emissions are estimated at 38.0 Gt $CO_2$ in 2026, a change of −34.1 Mt (−0.1%) from 2025. Regional increases of 0.331 Gt and decreases of 0.365 Gt largely offset one another. China’s reductions are concentrated in industry and ground transport. India’s growth is led by power and industry. The hindcast-adjusted range encompasses either direction of change. The regional results and event windows show the timing and composition of changes that are muted in the annual total.

**Acknowledgements**

None.

**Author contribution statement**

Tao Li contributed to conceptualization, methodology, data curation, formal analysis, visualization, drafting the manuscript, and reviewing and editing it; Zhu Deng contributed to conceptualization, methodology, data curation, manuscript review and editing, and supervision; Philippe Ciais contributed to conceptualization, methodology, manuscript review and editing, and supervision; Steve J. Davis contributed to interpretation and manuscript review and editing;

Ling Ye contributed to manuscript review and editing; Zhu Liu contributed to conceptualization, manuscript review and editing, and supervision.

**Data and code availability**

The Carbon Monitor activity estimates are available from https://carbonmonitor.org/. The daily projections and analysis code for this study are available from the corresponding authors upon request.